# An algorithm for computing Gröbner basis and the complexity evaluation

Yong-Jin Kim[1,2], Hyon-Song Paek[2], Nam-Chol Kim[3],*, Chong-Il Byon[4]

[1] Natural Science Academy, Kim Il Sung University, Pyongyang, DPR of Korea
[2] Faculty of Mathematics, Kim Il Sung University, Pyongyang, DPR of Korea
[3] Faculty of Physics, Kim Il Sung University, Pyongyang, DPR of Korea
[4] College of Computer Science, Kim Il Sung University, Pyongyang, DPR of Korea

∗ Corresponding author e-mail address: ryongnam19@yahoo.com

**ABSTRACT**

The first Gröbner basis algorithm was constructed by Buchberger in 1965; thus it bears his name to this day – Buchberger's algorithm.[9] Though Buchberger's algorithm looks relatively simple, it can take a very large amount of time. The step that creates $h_0$ via a normal form calculation is computationally very difficult. This is particularly frustrating (and wasteful) if the normal form calculation results in $h_0 = 0$ because all that computation ends up adding nothing to the Gröbner basis ($h_0$ is only added if it is non-zero). It would be nice if there were some way to "see ahead" and eliminate critical pairs whose S-polynomials top-reduce to 0 rather than actually computing the normal form of those S-polynomials and getting 0. The person who find this method is Faugère.

In 2002, J.C. Faugère published an algorithm called F5 in [3]. This algorithm has been shown in empirical tests to be the fastest Gröbner-basis-generating algorithm devised. But this original version of F5 is given in programming codes, so it is a bit difficult to understand. So By Yao Sun and Dingkang Wang, the F5 algorithm is simplified as F5B in a Buchberger's style such that it is easy to understand and implement in 2010.[10] And in 2008 Justin Gash supposed F5t that is the advance of F5 in [6], in 1999 J.C. Faugere supposed F4 in [2], so many algorithms to construct Grobner bases suggested.

But with this, it is the important problem to construct an efficient algorithm which can be done F5-reduction more quickly, too.

In this paper, we suggest a new efficient algorithm in order to compute S-polynomial reduction rapidly in the known algorithm for computing Gröbner bases, and compare the complexity with others.

**KEYWORDS**

Gröbner basis; F5 algorithm; Buchberger algorithm

## 1. INTRODUCTION

First we introduce some algorithms for computing Gröbner basis.

Let $F$ be a finite subset of $K[\underline{X}]$, then we can compute Gröbner basis $G$ in $K[\underline{X}]$ such that $F \subseteq G, Id(G) = Id(F)$ as follows.

### 1.1 Buchberger algorithm[1,2,4,6,8,9]

Given: $F$ = a finite subset of $K[\underline{X}]$

Find: $G$ = a Gröbner basis in $K[\underline{X}]$ with $F \subseteq G$ and $Id(G) = Id(F)$

Begin: $G \leftarrow F$

    $B \leftarrow \{\{g_i, g_j\} \mid g_i, g_j \in G, g_i \neq g_j\}$

    While $B \neq \phi$ do

        select $\{g_1, g_2\}$ from $B$

        $B \leftarrow B \setminus \{\{g_1, g_2\}\}$

$$h \leftarrow spol(g_1, g_2)$$

$$h_0 \leftarrow \text{some normal form of } h \text{ modulo } G$$

If $h_0 \neq 0$ then

$$B \leftarrow B \cup \{\{g, h_0\} \mid g \in G\}$$

$$G \leftarrow G \cup \{h_0\}$$

End if

End while

End

This algorithm takes a very large amount of time to reduce S-polynomial to 0.

To avoid this, many algorithms have been presented. F5B algorithm presented recently in [10] is one of the most efficient algorithms.

## 1.2 F5B algorithm [2, 3, 5, 6, 10, 11]

Input: $f_i \in K[\underline{X}], i = \overline{1, m}$

Output: The Gröbner basis $G$ of $\text{Id}(f_1, \cdots, f_m)$

Begin:

$$F_i := (e_i, f_i), i = \overline{1, m}$$

$$B := \{F_i \mid i = \overline{1, m}\}$$

$$CP := \{[F_i, F_j] \mid 1 \leq i < j \leq m\}$$

While $CP \neq \phi$ do

$cp :=$ an element of $CP$

$CP := CP \setminus \{cp\}$

If ($cp$ meets neither Syzygy Criterion nor Rewritten Criterion) then

$sp :=$ S-polynomial of $cp$

$P :=$ the reduction result of $sp$ by $B$

If $poly(P) \neq 0$ then

$$CP := CP \cup \{[P, Q] \mid Q \in B\}$$

End if

$B := B \cup \{P\}$

End if

End while

Return $\{poly(Q) \mid Q \in B\}$

End

The following is a subalgorithm of F5B algorithm to obtain the reduction result by $B$

## 1.3 Reduction algorithm[3, 5, 6, 10, 11]

Input: a set of signed polynomials $Todo$ which need to be reduced, a set of signed polynomials $B$

Output: a set of signed polynomials $Done$ in normal form by the set $B$.

Begin:

    $Done := \phi$

    While $Todo \neq \phi$ do

        $F :=$ the signed polynomial with minimal signature in set $Todo$

        $Todo := Todo \setminus \{F\}$

        $(Done', Todo') := F5-\text{reduction}(F, B, Todo)$

        $Done := Done \cup Done'$

        $Todo := Todo \cup Todo'$

        return $Done$

    End while

End

The following is F5-reduction algorithm of above algorithm.

## 1.4 F5-reduction algorithm[10]

Input: a signed polynomial $F$, a set of signed polynomials $B$

output: a 2-tuple $(Done, Todo)$

Begin:

    $G :=$ a signed polynomial in $B$ such that

        (1) $HT(G) \mid HT(F)$ and denote $v := \dfrac{HM(F)}{HM(G)}$

        (2) signature $sign(F) > sign(v \cdot G)$

        (3) $v \cdot G$ is not divisible by $B$, and

        (4) $v \cdot G$ is not rewritable by $B$

    If such $G$ does not exist then

        return$(\{F\}, \phi)$

    Else

        return$(\phi, \{F - vG\})$

    End if

End

    So we put the research results of previous papers together, and then we will obtain the following conclusion conclusion: by this time, the research has been done which find the S-polynomials that can be reduced to 0 by using Syzygy Criterion and Rewrittern Criterion. Especially in F5B algorithm it is the important problem to construct an efficient algorithm which can be done F5-reduction more quickly.

In this paper we will establish the following problem such as:

**First:** we are going to suggest an efficient algorithm which can be done S-polynomial reduction quickly in the algorithms for computing Gröbner basis presented by this time.

**Second:** we are going to compare the complexity of the suggested algorithm and the previous algorithms for computing Gröbner basis.

## 2. AN ALGORITHM FOR COMPUTING GRÖBNER BASIS

In this section we focus our attention on making a pair-selection-method newly, and by using it constructing an algorithm for Gröbner basis updated the reduction sub-algorithm of F5B algorithm.

Let polynomial $h$ need to be reduced by a set of polynomials $G$. Here

$$h = a_1 X^{h_1} + a_2 X^{h_2} + \cdots + a_p X^{h_p},$$
$$f = b_1 X^{f_1} + b_2 X^{f_2} + \cdots + b_q X^{f_q} \in G.$$

When $h$ is reduced by $f$, if $u = \dfrac{X^{h_1}}{X^{f_1}}$ then one reduction step of $h$ by $f$ is

$$h - \frac{a_1}{b_1} uf = (a_1 X^{h_1} + a_2 X^{h_2} + \cdots + a_p X^{h_p}) -$$
$$- \frac{a_1}{b_1}(ub_1 X^{f_1} + ub_2 X^{f_2} + \cdots + ub_q X^{f_q}) =$$
$$= a_2 X^{h_2} + \cdots + a_p X^{h_p} - \frac{a_1}{b_1} b_2 u X^{f_2} - \cdots \frac{a_1}{b_1} b_q u X^{f_q}$$

In every reduction step, the deg of $HT(h)$ becomes lower as fast as possible, $h$ will be reduced through the most rapidly reduction step.

If $h - uf = h'$ then $HT(h') = \max\{X^{h_2}, uX^{f_2}\}$, as the deg of $uX^{f_2}$ becomes lower, $h$ is reduced faster.

There are two cases with $X^{h_2}$ and $uX^{f_2}$.

Case1: $X^{h_2} > uX^{f_2}$

In this case $HT(h') = X^{h_2}$, so in the next reduction step $X^{h_2}$ is removed. And then let $uX^{f_2}$ become a head term through the many reduction steps. From this, as the deg of $uX^{f_2}$ is low, it is reduced faster.

Case2: $X^{h_2} \leq uX^{f_2}$

In this case $HT(h') = uX^{f_2}$, as the deg of $uX^{f_2}$ is low, it is reduced faster. So we can know that the deg of $uX^{f_2}$ is as low as possible to be reduced rapidly. So let polynomial $h = a_1 X^{h_1} + a_2 X^{h_2} + \cdots + a_p X^{h_p}$ need to be reduced by a set $G$. If $h$ is reduced by $f_k$ satisfying the following conditions, then S-polynomial reduction has been done more quickly than F5B algorithm.

$$f_i = b_{i_1} X^{i_1} + b_{i_2} X^{i_2} + \cdots + b_{i_q} X^{i_q} \; (i = \overline{1,m}) \in G$$

(1) $u_k = \dfrac{HT(h)}{HT(f_k)} \in T$

(2) $u_k \cdot X^{k_2} = \min_{<}\{u_i \cdot X^{i_2}\}$

Upon this we will construct the algorithm for computing Gröbner basis as following.

## 2.1 S-polynomial reduction algorithm

Input: a signed polynomial $sp \in K[X]$, a set of signed polynomials $B = \{F_1, \cdots, F_m\}$

Output: a signed polynomial $sp_0$ such that $sp \xrightarrow[B]{*} sp_0$

Begin:

    $h := poly(sp);$

    $f_i := poly(F_i)(i = \overline{1, m})$

    $G := \{f_1, \cdots, f_m\}$

    $f_i :=$ an element of $G$

    $t_{i1} := HM(f_i)$

    $t_{i2} := \max(T(f_i) \setminus \{HT(f_i)\})$

    $k := 1$

    While $k \neq 0$ do

        $k :=$ reduction sequence algorithm $(h, G)$

        if $k = 0$ then

            return $sp$

        else

$$u := \frac{HT(h)}{HT(f_k)}$$

            $sp := sp - u \cdot F_k$

            $h := poly(sp)$

        End if

    End while

Return $sp$

Following is the reduction sequence algorithm of above algorithm.

## 2.2 Reduction sequence algorithm $(h, G)$

Input: polynomial $h \in K[X], G = \{f_1, \cdots, f_m\}$

Output: index $k$ of $f_k$ which is chosen to reduce $h$ rapidly

Begin:

    $temp1 = 0, temp2 = HT(h);$

    For $i = 1$ to $|G|$

        If $HT(f_i) | HT(h)$ then

$$u := \frac{HT(h)}{HT(f_i)}$$

```
            If u·t_{i2} < temp2 then
                    temp2 = u·t_{i2}
                    temp1 = i
            End if
        End if
    End for
Return temp1
```

## 2.3 Correctness of algorithm

The algorithm can stop when it doesn't satisfy the condition of the while-loop. In other words $CP = \phi$.

The density of $B$ isn't over the number of all terms as possible at most. So after that the density of $B$ doesn't increase any more and on the contrary the density of $CP$ decrease continuously. So algorithm stops exactly.

When algorithm stops, let a set of signed polynomials be $B$. Then we know that S-polynomial reduction $(spol(F,G), B) = 0$ for arbitrary signed polynomial $F, G \in B$ from S-polynomial reduction algorithm. So the set of polynomials $\{poly(Q) | Q \in B\}$ becomes a Gröbner basis of $\text{Id}(f_1, \cdots, f_m) \subset K[X]$.

## 3. COMPLEXITY COMPARITION AND EVALUATION

In this section we compare the complexities of Buchberger algorithm, F5B algorithm and the algorithm that we suggest. The complexity evaluation is done following the algorithm.[7]

**Input of algorithm**: $F = \{f_1, \cdots, f_m\} \subseteq K[X_1, \cdots, X_n]$

**Output of algorithm**: Gröbner basis $G$ of $\text{Id}(f_1, \cdots, f_m)$

**Compiexity evaluation measure**: an operation in the ground field is counted as one step.

**Goal**: Calculation the number of steps [9]

$D := \max \deg(h_i | h_i \in H), H$ : a set of occuring polynomials during the computation

$$D \leq (8 \cdot \max \deg(F) + 1) \cdot 2^{\mindeg(F)} \quad [9]$$

Here $\max \deg(F)$ means maximum in the degs of polynomials of $F$, $\min \deg(F)$ means minimum.

The number of terms of n-variables, D-degree polynomial $f$ is

$$N(D,n) = \sum_{k=0}^{D} \binom{n+k-1}{k} = \sum_{k=0}^{D} \binom{n+k-1}{n-1} = \binom{n+D-1+1}{n-1+1} = \binom{n+D}{n}.$$

## 3.1 Complexity of Buchberger algorithm

**Proposition 1.** The complexity of Buchberger algorithm is

$$\frac{3}{2}n \cdot N(D,n)^5 + N(D,n)^4 + (2mn - \frac{1}{2}n + 7) \cdot N(D,n)^3 +$$
$$+ (-m^2 n - \frac{3}{2}n - \frac{3}{2}) \cdot N(D,n)^2 + (-\frac{1}{2}m^2 n - mn - \frac{3}{2}n - \frac{1}{2}) \cdot N(D,n)$$

**Proof:**

Begin:

$G := \{f_1, \cdots, f_m\}$

$B := \{\{g_1, g_2\}, \cdots, \{g_{m-1}, g_m\}\}$

While $B \neq \phi$ do

Lets begin our discussion by focusing our attention to $(i+1)$ th loop.

- select $\{g_a, g_b\}$ from $B_i$

- $B_i := B_i \setminus \{g_a, g_b\}$

- The complexity of " $h = spol(g_a, g_b)$ "

    • The complexity of finding the head term $t_a = HT(g_a)$ in $N(D,n)$ terms of $g_a$ is $2n \cdot N(D,n)$.

    • The complexity of finding the head term $t_b = HT(g_b)$ in $N(D,n)$ terms of $g_b$ is $2n \cdot N(D,n)$.

    • The complexity of computing $lcm(t_a, t_b) = t$ is $n$.

    • The complexity of computing $s_a$ such as $t = s_a t_a$ is $n$.

    • The complexity of computing $s_b$ such as $t = s_b t_b$ is $n$.

    • The complexity of computing $spol(g_a, g_b) = c_b s_a g_a - c_a s_b g_b$ is $(2n+3) \cdot N(D,n) + n \cdot N(D,n)^2$.

        ✓ The complexity of computing $c_b s_a g_a$ is $(n+1) \cdot N(D,n)$.

        ✓ The complexity of computing $c_a s_b g_b$ is $(n+1) \cdot N(D,n)$.

        ✓ The complexity of computing $c_b s_a g_a - c_a s_b g_b$ is $n \cdot N(D,n)^2 + N(D,n)$.

    So the complexity of computing $h = spol(g_a, g_b)$ is

    $$n \cdot N(D,n)^2 + (6n+3) \cdot N(D,n) + 3n \quad (1)$$

- The complexity of computing $h \xrightarrow[G]{*} h_0$

    • The complexity of computing $HT(g)$ for any $g$ in $G_i$ is $2n \cdot N(D,n) \cdot |G_i|$

    • The complexity of computing $HT(h)$ is $2n \cdot N(D,n)$

    • The complexity of computing $g$ such as $HT(g)|HT(h)$ in $G_i$ is $n \cdot |G_i|$.

    • The complexity of computing $v_{k_1}$ such as $HT(h) = v_{k_1} \cdot HT(g_{k_1})$ is $n$.

- The complexity of computing $h \xrightarrow{G} h - \dfrac{HC(h)}{HC(g_{k_1})} v_{k_1} g_{k_1} = h_1$ is

   $n \cdot N(D,n)^2 + (n+2) \cdot N(D,n) + 1$

   ✓ The complexity of computing $\dfrac{HC(h)}{HC(g_{k_1})} v_{k_1} g_{k_1}$ is $1 + (1+n) \cdot N(D,n)$.

   ✓ The complexity of computing $h - \dfrac{HC(h)}{HC(g_{k_1})} v_{k_1} g_{k_1}$ is $n \cdot N(D,n)^2 + N(D,n)$.

   … … …

   So the complexity of computing $h \xrightarrow{*}_{G} h_0$ is

$$N(D,n) \cdot (n \cdot N(D,n)^2 + (2n \cdot N(D,n) + n) \cdot |G_i| + (3n+2) \cdot N(D,n) + n + 1) = \\ = n \cdot N(D,n)^3 + (2n \cdot N(D,n)^2 + n \cdot N(D,n)) \cdot |G_i| + (3n+2) \cdot N(D,n)^2 + (n+1) \cdot N(D,n) \quad (2)$$

- The complexity of computing "If $h_0 \neq 0$ then" is 1. (3)

   • $B_{i+1} = B_i \cup \{\{g, h_0\} \mid g \in G_i\}$

   • $G_{i+1} = G_i \cup \{h_0\}$

   End if

   So the complexity of computing $(i+1)$th loop is

$$T_i = (1) + (2) + (3) = n \cdot N(D,n)^3 + (2n \cdot N(D,n)^2 + n \cdot N(D,n)) \cdot |G_i| + \\ + (4n+2) \cdot N(D,n)^2 + (7n+4) \cdot N(D,n) + 3n + 1$$

End while

So all complexity is $T = \sum\limits_{i=1}^{W} T_i$.

Here $W$ is the number of while-loop of algorithm.

End.

Now let's compute $|G_i|$ and $W$.

$|G_0| = m$, $|B_0| = \binom{m}{2} = \dfrac{m(m-1)}{2}$ and the number of while-loop is $W = (N(D,n) - m) + |B_{N(D,n)-m}|$.

For $i$, such as $1 \leq i \leq (N(D,n) - m) + |B_{N(D,n)-m}|$, we will go through in two cases as following.

**Case1**: $1 \leq i \leq N(D,n) - m$

$$\begin{cases} |G_i| = m + i \\ |B_0| = \dfrac{m(m-1)}{2} \\ |B_{i+1}| = |B_i| - 1 + |G_i| = |B_i| + (m+i-1) \end{cases}$$

Let $A(x) = \sum_{i \geq 0} B_i \cdot x^i$.

Let's multiply both sides by $x^i$ and sum up both sides.

$$\sum_{i \geq 0} B_{i+1} \cdot x^i = \sum_{i \geq 0} B_i \cdot x^i + \sum_{i \geq 0} i \cdot x^i + (m-1) \sum_{i \geq 0} x^i \cdot$$

$$\cdot \frac{1}{x}(A(x) - A(0)) = A(x) + \frac{x}{(1-x)^2} + (m-1)\frac{1}{1-x} \cdot \left(\frac{1}{x} - 1\right) A(x) = \frac{1}{x} \frac{m(m-1)}{2} + \frac{x}{(1-x)^2} + (m-1)\frac{1}{1-x}$$

$$A(x) = \frac{m(m-1)}{2} \frac{1}{1-x} + \frac{x^2}{(1-x)^3} + (m-1)\frac{x}{(1-x)^2} =$$

$$= \frac{m(m-1)}{2} \sum_{i \geq 0} x^i + \sum_{i \geq 0} \frac{i(i-1)}{2} x^i + \sum_{i \geq 0} (m-1) i x^i =$$

$$= \sum_{i \geq 0} \left( \frac{m(m-1)}{2} + (m-1)i + \frac{i(i-1)}{2} \right) x^i$$

$$\therefore |B_i| = \frac{m(m-1)}{2} + (m-1)i + \frac{i(i-1)}{2}$$

$$\therefore |B_{N(D,n)-m}| = \frac{(N(D,n) - m - 1)(N(D,n) - m)}{2} + (m-1)(N(D,n) - m) + \frac{m(m-1)}{2}$$

$$= \frac{1}{2} N(D,n)^2 - \frac{3}{2} N(D,n) + m$$

Case2: $N(D,n) - m + 1 \leq i \leq W$

$$|G_i| = N(D,n)$$

$$|B_i| = |B_{i-1}| - 1$$

$$\therefore T = \sum_{i=1}^{W} T_i = \sum_{i=1}^{N(D,n)-m} T_i + \sum_{i=N(D,n)-m+1}^{W} T_i =$$

$$= \sum_{i=1}^{N(D,n)-m} [n \cdot N(D,n)^3 + (2n \cdot N(D,n)^2 + n \cdot N(D,n)) \cdot |G_i|$$

$$+ (4n+2) \cdot N(D,n)^2 + (7n+4) \cdot N(D,n) + 3n + 1]$$

$$+ |B_{N(D,n)-m}| \cdot [n \cdot N(D,n)^3 + (2n \cdot N(D,n)^2 + n \cdot N(D,n)) \cdot |G_i|$$

$$+ (4n+2) \cdot N(D,n)^2 + (7n+4) \cdot N(D,n) + 3n + 1] =$$

$$= \frac{3}{2} n \cdot N(D,n)^5 + N(D,n)^4 + (2mn - \frac{1}{2}n + 7) \cdot N(D,n)^3$$

$$+ (-m^2 n - \frac{3}{2} n - \frac{3}{2}) \cdot N(D,n)^2 + (-\frac{1}{2} m^2 n - mn - \frac{3}{2} n - \frac{1}{2}) \cdot N(D,n)$$

So the complexity of Buchberger algorithm is

$$T = \frac{3}{2} n \cdot N(D,n)^5 + N(D,n)^4 + (2mn - \frac{1}{2} n + 7) \cdot N(D,n)^3 +$$

$$+ (-m^2 n - \frac{3}{2} n - \frac{3}{2}) \cdot N(D,n)^2 + (-\frac{1}{2} m^2 n - mn - \frac{3}{2} n - \frac{1}{2}) \cdot N(D,n)$$

This completes the proof.   □

## 3.2 Complexity of F5B algorithm

**Proposition 2.** The complexity of F5B algorithm is

$$\frac{2n}{3} \cdot N(D,n)^5 + (mn + \frac{14}{3}n + \frac{2}{3}) \cdot N(D,n)^4 + (-m^2n + mn + m + \frac{11}{6}n + 2) \cdot N(D,n)^3 +$$

$$+ (\frac{10}{3}m^3n - 3m^2n - m^2 + \frac{14}{3}mn + 2m - \frac{16}{3}n - \frac{2}{3}) \cdot N(D,n)^2 +$$

$$+ (-\frac{2}{3}m^3n - \frac{2}{3}m^3 + m^2n + \frac{17}{6}mn + \frac{8}{3}m - \frac{11}{2}n - 2) \cdot N(D,n) +$$

$$+ (-\frac{11}{2}m^2n - 4m^2 + \frac{7}{2}mn + 2m)$$

**Proof:**

Begin:

- $i = \overline{1,m}, F_i = (e_i, f_i)$

- $B_0 = \{F_1, \cdots, F_m\}$

- $CP_0 := \{[F_i, F_j] \mid 1 \leq i < j \leq m\}$

  ($CP_0 = \{[F_1, F_2], \cdots, [F_{m-1}, F_m]\}$)

   • The complexity of computing $[F_i, F_j] = [u, F_i, v, F_j]$ is $4n \cdot N(D,n) + 3n$.

   ✓ The complexity of computing is $t_i = HT(poly(F_i))$ is $2n \cdot N(D,n)$.

   ✓ The complexity of computing is $t_j = HT(poly(F_j))$ is $2n \cdot N(D,n)$.

   ✓ The complexity of computing $lcm(t_i, t_j) = t$ is $n$.

   ✓ The complexity of computing $u$ such as $t = ut_i$ is $n$.

   ✓ The complexity of computing $v$ such as $t = vt_j$ is $n$.

   So the complexity of computing $CP_0$ is

$$\frac{m(m-1)}{2} \cdot (4n \cdot N(D,n) + 3n) \quad (4)$$

- While $CP_i \neq \phi$ do

   Let's begin our discussion by focusing our attention to $i$ − th loop.

   - $cp \leftarrow CP_{i-1}$
   - $CP_i \leftarrow CP_{i-1} \setminus \{cp\}$
   - If $cp$ meets neither Syzygy Criterion nor Rewritten Criterion, then

      • $cp$ meets Syzygy Criterion?

      $cp = (u, F, v, G)$

      ✓ $uF$ is divisible?

         The complexity of computing $u \cdot S(F)$ is $n$

         The complexity of computing $HT(g)$ is $2n \cdot N(D,n)$.

The complexity of computing $HT(g)|u \cdot S(F)$ is $n$.

The complexity of computing $i < j$ is 1

So for any $g$ in $B_{i-1}$, the complexity is $(2n \cdot N(D,n)+n+1) \cdot |B_{i-1}|$.

The complexity of computing "$uF$ is divisible?" is $(2n \cdot N(D,n)+n+1) \cdot |B_{i-1}|+n$.

✓ The complexity of computing "$vG$: is divisible?" is $(2n \cdot N(D,n)+n+1) \cdot |B_{i-1}|+n$.

- So the complexity of computing "$cp$: meets Syzygy Criterion?" is $2((2n \cdot N(D,n)+n+1) \cdot |B_{i-1}|+n)$.

- $cp$ meets Rewritten Criterion?

  ✓ $uF$ is rewritable?

  The complexity of computing $u \cdot S(F)$ is $n$.

  The complexity of computing $S(G)|u \cdot S(F)$ for $G$ in $G$ is $n$.

  The complexity of computing "$G$ is generated later than $F$?" is 1.

  So for any $g$ in $B_{i-1}$, the complexity is $(n+1) \cdot |B_{i-1}|$.

  So the complexity of computing "$uF$ is rewritable?" is $n+(n+1) \cdot |B_{i-1}|$.

  ✓ The complexity of computing "$vG$ is rewritable?" is $n+(n+1) \cdot |B_{i-1}|$.

The complexity of computing "$cp$ meets Rewritten Criterion?" is $2(n+(n+1) \cdot |B_{i-1}|)$.

So the complexity of computing "if-condition satisfy?" is

$$\begin{aligned}&2((2n \cdot N(D,n)+n+1) \cdot |B_{i-1}|+n) + 2(n+(n+1) \cdot |B_{i-1}|) = \\ &= 2((2n \cdot N(D,n)+2n+2) \cdot |B_{i-1}|+2n) = (4n \cdot N(D,n)+4n+4) \cdot |B_{i-1}|+4n\end{aligned} \quad (5)$$

then

- $sp \leftarrow$ S-polynomial of $cp$

  $cp = (u, F, v, G) = (u,(e_i, f), v,(e_j, g))$

  $spol(cp) = HC(g)uF - HC(f)vG = (HC(g)ue_i, HC(g)uf) - (HC(f)ve_j, HC(f)vg)$

  ✓ $HC(g)uF = (HC(g)ue_i, HC(g)uf) : (n+1)m+(n+1) \cdot N(D,n)$

  ✓ $HC(f)vG = (HC(f)ve_j, HC(f)vg) : (n+1)m+(n+1) \cdot N(D,n)$

  $sp = HC(g)uF - HC(f)vG =$

  ✓ $= (HC(g)ue_i, HC(g)uf) - (HC(f)ve_j, H(f)vg) : (n \cdot N(D,n)^2 + N(D,n)) \cdot m +$
  $+ n \cdot N(D,n)^2 + N(D,n)$

  The complexity of computing $sp$ is

$$\begin{aligned}&2((n+1)m+(n+1) \cdot N(D,n)) + (n \cdot N(D,n)^2 + N(D,n)) \cdot m + \\ &+ n \cdot N(D,n)^2 + N(D,n) = (mn+n) \cdot N(D,n)^2 + (2n+m+3) \cdot N(D,n) + \\ &+ 2(n+1)m\end{aligned} \quad (6)$$

- $p \leftarrow$ reduction result of $sp$ ( $sp \xrightarrow[B_{i-1}]{*} p$ )

- ✓ $Done \leftarrow \phi$;
- ✓ While $Todo \neq \phi$ do

    The complexity of computing "$F \leftarrow$ the signed polynomial with minimal signature in set $Todo$" is $|B_{i-1}| \cdot n$.

    $Todo \leftarrow Todo \setminus \{F\}$

    $(Done', Todo') \leftarrow F5-\text{reduction}(F, B_{i-1})$

    $G \leftarrow B_{i-1}$

    (1) The complexity of computing "$HT(G) | HT(F), v = \dfrac{HM(f)}{HM(G)}$" is $4n \cdot N(D,n) + 2n$.

     The complexity of computing "$HT(F)$" is $2n \cdot N(D,n)$.

     The complexity of computing "$HT(G)$" is $2n \cdot N(D,n)$.

     The complexity of computing "$HT(G) | HT(F)$" is $n$.

     The complexity of computing "$v = \dfrac{HM(f)}{HM(G)}$" is $n$.

    (2) The complexity of computing "$S(F) > S(v \cdot G)$" is $2n$

     The complexity of computing "$S(v \cdot G)$" is $n$

     The complexity of computing "$S(F) > S(v \cdot G)$" is $n$

    (3) The complexity of computing "$v \cdot G$ is not divisible by $B_{i-1}$" is $(2n \cdot N(D,n) + n + 1) \cdot |B_{i-1}| + n$.

    (4) The complexity of computing "$v \cdot G$ is not rewritable by $B_{i-1}$" is $n + (n+1) \cdot |B_{i-1}|$.

    $\therefore \{4n \cdot N(D,n) + 2n + 2n + (2n \cdot N(D,n) + n + 1) \cdot |B_{i-1}| + n + n + (n+1) \cdot |B_{i-1}|\} \cdot B_{i-1} =$
    $= (2n \cdot N(D,n) + 2n + 2) \cdot |B_{i-1}|^2 + 4n \cdot N(D,n) \cdot |B_{i-1}| + 6n \cdot |B_{i-1}|$

    If such $G$ does not exist

    then return $(\{F\}, \phi)$

    else return $(\phi, \{F - v \cdot G\})$:

    $mn \cdot N(D,n) + (n \cdot N(D,n)^2 + N(D,n)) \cdot m + n \cdot N(D,n)^2 + N(D,n)$

     The complexity of computing $v \cdot G$ is $mn \cdot N(D,n)$.

     The complexity of computing $F - v \cdot G$ is

     $(n \cdot N(D,n)^2 + N(D,n)) \cdot m + n \cdot N(D,n)^2 + N(D,n)$.

    End if

    $Done \leftarrow Done \cup Done'$

    $Todo \leftarrow Todo \cup Todo'$

- ✓ End while
- ✓ return $Done$

The complexity of F5-reduction algorithm is

$$\{|B_{i-1}| \cdot n + (2n \cdot N(D,n) + 2n + 2) \cdot |B_{i-1}|^2 + 4n \cdot N(D,n) \cdot |B_{i-1}| + 6n \cdot |B_{i-1}| + mn \cdot N(D,n) +$$
$$+ (n \cdot N(D,n)^2 + N(D,n)) \cdot m + n \cdot N(D,n)^2 + N(D,n)\} \cdot N(D,n) =$$
$$= (2n \cdot N(D,n)^2 + (2n+2) \cdot N(D,n)) \cdot |B_{i-1}|^2 + (4n \cdot N(D,n)^2 + 7n \cdot N(D,n)) \cdot |B_{i-1}| +$$
$$+ (mn+n) \cdot N(D,n)^3 + (mn+m+1)N(D,n)^2 \tag{7}$$

- If $poly(P) \neq 0$ then

  The complexity of computing "$CP_i \leftarrow CP_i \cup \{[P,Q] | Q \in B_{i-1}\}$" is

  $$(4n \cdot N(D,n) + 3n) \cdot |B_{i-1}|. \tag{8}$$

  Endif

- $B_i \leftarrow B_{i-1} \cup \{P\}$

  - Endif

  The complexity of $i$-th loop is

$$T_i = (5) + (6) + (7) + (8)$$
$$= (4n \cdot N(D,n) + 4n + 4) \cdot |B_{i-1}| + 4n + (mn+n) \cdot N(D,n)^2 +$$
$$+ (2n+m+3) \cdot N(D,n) + 2(n+1)m + (2n \cdot N(D,n)^2 +$$
$$+ (2n+2) \cdot N(D,n)) \cdot |B_{i-1}|^2 + (4n \cdot N(D,n)^2 + 7n \cdot N(D,n)) \cdot |B_{i-1}| + (mn+n) \cdot N(D,n)^3 +$$
$$+ (mn+m+1) \cdot N(D,n)^2 + (4n \cdot N(D,n) + 3n) \cdot |B_{i-1}| =$$
$$= (2n \cdot N(D,n)^2 + (2n+2) \cdot N(D,n)) \cdot |B_{i-1}|^2 + (4n \cdot N(D,n)^2 + 15n \cdot N(D,n) + 7n+4) \cdot |B_{i-1}| +$$
$$+ (mn+n) \cdot N(D,n)^3 + (2mn+m+n+1) \cdot N(D,n)^2 + (2n+m+3) \cdot N(D,n) + 4n + 2(n+1)m$$

- End while

- return $\{poly(Q) | Q \in B_i\}$

$$T = (4) + \sum_{i=1}^{W} T_i = \frac{m(m-1)}{2} \cdot (4n \cdot N(D,n) + 3n) + \sum_{i=1}^{W} T_i$$

End

Now let's compute $|G_i|$ and $W$.

$$\begin{cases} |B_0| = m \\ |CP_0| = \frac{m(m-1)}{2} \\ |B_i| = |B_{i-1}| + 1 = m + i \\ |CP_i| = |CP_{i-1}| - 1 + |B_{i-1}| = |CP_{i-1}| - 1 + m + i - 1 \end{cases}$$

$$|CP_i| = \frac{i(i-1)}{2} + (m-1)i + \frac{m(m-1)}{2}, 1 \leq i \leq W$$

The number of while-loop is $W = (N(D,n) - m) + |CP_{N(D,n)-m}|$.

For $i$ such as $1 \leq i \leq (N(D,n) - m) + |B_{N(D,n)-m}|$ we will go through in two cases as following.

Case1: $1 \leq i \leq N(D,n) - m$, $|B_{i-1}| = m + i - 1$

$$T_i = (2n \cdot N(D,n)^2 + (2n+2) \cdot N(D,n)) \cdot |B_{i-1}|^2 + (4n \cdot N(D,n)^2 + 15n \cdot N(D,n) + 7n + 4) \cdot |B_{i-1}| +$$
$$+ (mn+n) \cdot N(D,n)^3 + (2mn+m+n+1) \cdot N(D,n)^2 + (2n+m+3) \cdot N(D,n) + 4n + 2(n+1)m =$$
$$= (2n \cdot N(D,n)^2 + (2n+2) \cdot N(D,n)) \cdot i^2 +$$
$$+ (4mn \cdot N(D,n)^2 + (4mn+4m+11n-4) \cdot N(D,n) + 7n + 4)i$$
$$+ (mn+n) \cdot N(D,n)^3 + (2m^2n + 2mn + m - n + 1) \cdot N(D,n)^2$$
$$+ (2m^2n + 11mn + 2m^2 - 11n - 3m + 5) \cdot N(D,n) + 9mn + 6m - 3n - 4$$

$$\sum_{i=1}^{N(D,n)-m} T_i$$
$$= (2n \cdot N(D,n)^2 + (2n+2) \cdot N(D,n)) \cdot \sum_{i=1}^{N(D,n)-m} i^2 + (4mn \cdot N(D,n)^2 +$$
$$+ (4mn + 4m + 11n - 4) \cdot N(D,n) + 7n + 4) \cdot \sum_{i=1}^{N(D,n)-m} i +$$
$$+ ((mn+n) \cdot N(D,n)^3 + (2m^2n + 2mn + m - n + 1) \cdot N(D,n)^2$$
$$+ (2m^2n + 11mn + 2m^2 - 11n - 3m + 5) \cdot N(D,n) +$$
$$+ 9mn + 6m - 3n - 4) \cdot (N(D,n) - m) =$$
$$= \frac{2n}{3} \cdot N(D,n)^5 + (mn + \frac{8}{3}n + \frac{2}{3}) \cdot N(D,n)^4 + (-m^2n + mn + m + \frac{35}{6}n)N(D,n)^3 +$$
$$+ (\frac{10}{3}m^3n - 3m^2n - m^2 + \frac{2}{3}mn - \frac{4}{3}n + \frac{16}{3})N(D,n)^2 +$$
$$+ (-\frac{2}{3}m^3n - \frac{2}{3}m^3 - m^2n + \frac{5}{6}mn - \frac{4}{3}m + \frac{1}{2}n - 2)N(D,n) +$$
$$+ (-\frac{11}{2}m^2n - 4m^2 - \frac{1}{2}mn + 2m) \quad (9)$$

Case2: $N(D,n) - m + 1 \le i \le W$

$|B_{i-1}| = N(D,n)$

$T_i = (4n \cdot N(D,n) + 4n + 4) \cdot |B_{i-1}| + 4n = 4n \cdot N(D,n)^2 + (4n+4) \cdot N(D,n) + 4n$

$$\therefore \sum_{i=N(D,n)-m+1}^{W} T_i = |CP_{N(D,n)-m}| \cdot T_i$$
$$= [\frac{(N(D,n)-m)(N(D,n)-m+1)}{2} + (m-1)(N(D,n)-m) + \frac{m(m-1)}{2}] \cdot$$
$$\cdot (4n \cdot N(D,n)^2 + (4n+4) \cdot N(D,n) + 4n) \quad (10)$$
$$= 2n \cdot N(D,n)^4 - (4n-2) \cdot N(D,n)^3 + (4mn - 4n - 6) \cdot N(D,n)^2 +$$
$$+ (4mn + 4m - 6n) \cdot N(D,n) + 4mn$$

$$\therefore T = \frac{m(m-1)}{2} \cdot (4n \cdot N(D,n) + 3n) + (9) + (10)$$

$$= \frac{2n}{3} \cdot N(D,n)^5 + (mn + \frac{14}{3}n + \frac{2}{3}) \cdot N(D,n)^4 + (-m^2n + mn + m + \frac{11}{6}n + 2) \cdot N(D,n)^3$$

$$+ (\frac{10}{3}m^3n - 3m^2n - m^2 + \frac{14}{3}mn + 2m - \frac{16}{3}n - \frac{2}{3}) \cdot N(D,n)^2$$

$$+ (-\frac{2}{3}m^3n - \frac{2}{3}m^3 + m^2n + \frac{17}{6}mn + \frac{8}{3}m - \frac{11}{2}n - 2) \cdot N(D,n)$$

$$+ (-\frac{11}{2}m^2n - 4m^2 + \frac{7}{2}mn + 2m)$$

So the complexity of F5B algorithm is

$$T = \frac{2n}{3} \cdot N(D,n)^5 + (mn + \frac{14}{3}n + \frac{2}{3}) \cdot N(D,n)^4 + (-m^2n + mn + m + \frac{11}{6}n + 2) \cdot N(D,n)^3$$

$$+ (\frac{10}{3}m^3n - 3m^2n - m^2 + \frac{14}{3}mn + 2m - \frac{16}{3}n - \frac{2}{3}) \cdot N(D,n)^2$$

$$+ (-\frac{2}{3}m^3n - \frac{2}{3}m^3 + m^2n + \frac{17}{6}mn + \frac{8}{3}m - \frac{11}{2}n - 2) \cdot N(D,n) +$$

$$+ (-\frac{11}{2}m^2n - 4m^2 + \frac{7}{2}mn + 2m)$$

This completes the proof. □

### 3.3 Complexity of new algorithm

**Proposition 3.** The complexity of S-polynomial reduction algorithm is

$$(mn + 4n) \cdot N(D,n)^4 + (-m^2n + mn + m + \frac{15}{2}n + 31) \cdot N(D,n)^3$$

$$+ (-3m^2n - m^2 + 5mn - 5n - 1) \cdot N(D,n)^2$$

$$+ (-7m^2n + 10mn - m^2 - m - 2n - 2) \cdot N(D,n) + 4m^2n - 2m^2 + 2m$$

**Proof:**

▲ S-polynomial reduction algorithm

Input : a signed polynomial $sp \in K[X]$, a set of signed polynomials $B = \{F_1, \cdots, F_m\}$

Output : a signed polynomial $sp_0$ such that $sp \xrightarrow[B]{*} sp_0$

Begin:

- $h := poly(sp);$

- $f_i := poly(F_i)(i = \overline{1,m})$

- $G := (f_1, \cdots, f_m)$

  - $f_i :=$ an element of $G$
  - The complexity of computing "$t_{i1} := HM(f_i)$" is $2n \cdot N(D,n)$

- The complexity of computing is "$t_{i2} := \max(T(f_i) \setminus \{HT(f_i)\})$" is $2n \cdot N(D,n)$.
- $k := 1$

- The complexity of computing "While $k \neq 0$ do" is 1.

  - The complexity of computing "$k :=$ reduction sequence algorithm $(h, G)$" is $(2n \cdot N(D,n) + 3n) \cdot |B_{i-1}| + 2n \cdot N(D,n)$.
  - The complexity of computing "if $k = 0$ then" is 1.

    return $sp$.

    The complexity of computing "$u := \dfrac{HM(h)}{HM(f_k)}$" is $n$.

    The complexity of computing "$sp := sp - u \cdot F_k$" is

    $mn \cdot N(D,n) + (n \cdot N(D,n)^2 + N(D,n)) \cdot m + n \cdot N(D,n)^2 + N(D,n)$.

    $h := poly(sp)$

  Endwhile

  So the complexity of while-loop is

  $[(2n \cdot N(D,n) + 3n) \cdot |B_{i-1}| + (mn + n) \cdot N(D,n)^2 + (mn + m + 2n + 1) \cdot N(D,n) + n + 2] \cdot N(D,n)$.

- Return $sp$

So the complexity of S-polynomial reduction algorithm is

$$(2n \cdot N(D,n)^2 + 7n \cdot N(D,n)) \cdot |B_{i-1}| + (mn + n) \cdot N(D,n)^3 + \\ + (mn + m + 2n + 1) \cdot N(D,n)^2 + (n + 2) \cdot N(D,n) \tag{11}$$

▲ Reduction sequence algorithm $(h, G)$

**Input**: polynomial $h \in K[X], G = \{f_1, \cdots, f_m\}$

**Output**: index $k$ of $f_k$ which is chosen to reduce $h$ rapidly

Begin:

- $temp1 = 0$, The complexity of computing "$temp2 = HT(h)$" is $2n \cdot N(D,n)$.

- For $i = 1$ to $|G|$

  - The complexity of computing "If $HT(f_i) | HT(h)$ then" is $2n \cdot N(D,n) + n$.

    $u := \dfrac{HT(h)}{HT(f_i)}$

    The complexity of computing

    "If $u \cdot t_{i2} < temp2$ then" is $2n$.

    $temp2 = u \cdot t_{i2}$

    $temp1 = i$

    End if

    End if

  End for

  The complexity of for-while is $(2n \cdot N(D,n)+3n) \cdot |B_{i-1}|$.

 - Return $temp1$

End

So the complexity of reduction sequence algorithm is $(2n \cdot N(D,n)+3n) \cdot |B_{i-1}|+2n \cdot N(D,n)$.

When S-polynomial reduction algorithm is used instead of F5-reduction algorithm of F5B algorithm, it's complexity is as following.

The complexity of i-th loop is

$$T_i = (5)+(6)+(11)+(8)$$
$$= (4n \cdot N(D,n)+4n+4) \cdot |B_{i-1}|+4n++(mn+n) \cdot N(D,n)^2+(2n+m+3) \cdot N(D,n)+$$
$$+2(n+1)m+(2n \cdot N(D,n)^2+7n \cdot N(D,n)) \cdot |B_{i-1}|+(mn+n) \cdot N(D,n)^3+(mn+m+2n+1) \cdot N(D,n)^2+$$
$$+(n+2) \cdot N(D,n)+(4n \cdot N(D,n)+3n) \cdot |B_{i-1}|=$$
$$= (2n \cdot N(D,n)^2+15n \cdot N(D,n)+7n+4) \cdot |B_{i-1}|+(mn+n) \cdot N(D,n)^3+(2mn+m+3n+1) \cdot N(D,n)^2$$
$$+(3n+m+5) \cdot N(D,n)+2mn+6n$$

$$T = (4)+\sum_{i=1}^{W}T_i = \frac{m(m-1)}{2} \cdot (4n \cdot N(D,n)+3n)+\sum_{i=1}^{W}T_i$$

$$\begin{cases} |B_0|=m \\ |CP_0|=\dfrac{m(m-1)}{2} \\ |B_i|=|B_{i-1}|+1=m+i \\ |CP_i|=|CP_{i-1}|-1+|B_{i-1}|=|CP_{i-1}|-1+m+i-1 \\ |CP_i|=\dfrac{i(i-1)}{2}+(m-1)i+\dfrac{m(m-1)}{2}, 1 \le i \le W \\ W=(N(D,n)-m)+|CP_{N(D,n)-m}| \end{cases}$$

(1) $1 \le i \le N(D,n)-m$

$$|B_{i-1}|=m+i-1$$

$$T_i = (2n \cdot N(D,n)^2+15n \cdot N(D,n)+7n+4) \cdot |B_{i-1}|+(mn+n) \cdot N(D,n)^3+$$
$$+(2mn+m+3n+1) \cdot N(D,n)^2+(3n+m+5) \cdot N(D,n)+2mn+6n=$$
$$= (2n \cdot N(D,n)^2+15n \cdot N(D,n)+7n+4) \cdot (m+i-1)+(mn+n) \cdot N(D,n)^3+$$
$$+(2mn+m+3n+1) \cdot N(D,n)^2+(3n+m+5) \cdot N(D,n)+2mn+6n=$$
$$= (2n \cdot N(D,n)^2+15n \cdot N(D,n)+7n+4) \cdot i++(2n \cdot N(D,n)^2+15n \cdot N(D,n)+7n+4) \cdot (m-1)$$
$$+(mn+n) \cdot N(D,n)^3+(2mn+m+3n+1) \cdot N(D,n)^2+(3n+m+5) \cdot N(D,n)+2mn+6n=$$
$$=(2n \cdot N(D,n)^2+15n \cdot N(D,n)+7n+4) \cdot i+(mn+n) \cdot N(D,n)^3+(4mn+m+n+1) \cdot N(D,n)^2+$$
$$+(15mn-12n+m+3) \cdot N(D,n)+9mn+4m-n-4$$

$$\therefore \sum_{i=1}^{N(D,n)-m} T_i =$$

$$= (2n \cdot N(D,n)^2 + 15n \cdot N(D,n) + 7n + 4) \cdot \sum_{i=1}^{N(D,n)-m} i +$$

$$+ (N(D,n)-m)) \cdot ((mn+n) \cdot N(D,n)^3 + (4mn+m+n+1) \cdot N(D,n)^2$$

$$+ (15mn - 12n + m + 3) \cdot N(D,n) + 9mn + 4m - n - 4) =$$

$$= (2n \cdot N(D,n)^2 + 15n \cdot N(D,n) + 7n + 4) \cdot \frac{1}{2}(N(D,n)-m)(N(D,n)-m+1)$$

$$+ ((mn+n) \cdot N(D,n)^4 + (-m^2n + 3mn + m + n + 1) \cdot N(D,n)^3$$

$$+ (-4m^2n - m^2 + 14mn - 12n + 3) \cdot N(D,n)^2$$

$$+ (-15m^2n + 21mn - m^2 - m - n - 4) \cdot N(D,n) - 9m^2n - 4m^2 + mn + 4m) =$$

$$= (n \cdot N(D,n)^4 - (2mn - \frac{17}{2}n) \cdot N(D,n)^3 + (m^2n - 13mn + 11n + 2) \cdot N(D,n)^2 +$$

$$+ (6m^2n - 13mn - 4m + 5n + 2) \cdot N(D,n) + (\frac{7}{2}m^2n - \frac{7}{2}mn + 2m^2 - 2m)) +$$

$$+ ((mn+n) \cdot N(D,n)^4 + (-m^2n + 3mn + m + 3n + 1) \cdot N(D,n)^3 +$$

$$+ (-4m^2n - m^2 + 14mn - 12n + 3) \cdot N(D,n)^2 +$$

$$+ (-15m^2n + 21mn - m^2 - m - n - 4) \cdot N(D,n) - 9m^2n - 4m^2 + mn + 4m) =$$

$$= (mn + 2n) \cdot N(D,n)^4 + (-m^2n + mn + m + \frac{23}{2}n + 1) \cdot N(D,n)^3$$

$$+ (-3m^2n - m^2 + mn - n + 5) \cdot N(D,n)^2$$

$$+ (-9m^2n + 8mn - m^2 - 5m + 4n - 2) \cdot N(D,n) - \frac{11}{2}m^2n - 2m^2 - \frac{5}{2}mn + 2m$$

(12)

(2) $N(D,n) - m + 1 \leq i \leq W$

$$|B_{i-1}| = N(D,n)$$

$$T_i = (4n \cdot N(D,n) + 4n + 4) \cdot |B_{i-1}| + 4n =$$

$$= 4n \cdot N(D,n)^2 + (4n+4) \cdot N(D,n) + 4n$$

$$\therefore \sum_{i=N(D,n)-m+1}^{W} T_i = |CP_{N(D,n)-m}| \cdot T_i =$$

$$= [\frac{(N(D,n)-m)(N(D,n)-m-1)}{2} + (m-1)(N(D,n)-m) + \frac{m(m-1)}{2}] \cdot$$

$$\cdot (4n \cdot N(D,n)^2 + (4n+4) \cdot N(D,n) + 4n) =$$

$$= [\frac{1}{2}(N(D,n)^2 - 2m \cdot N(D,n) + m^2 - N(D,n) + m) +$$

$$+ m \cdot N(D,n) - N(D,n) - m^2 + m + \frac{1}{2}m^2 - \frac{1}{2}m] \cdot$$

$$\cdot (4n \cdot N(D,n)^2 + (4n+4) \cdot N(D,n) + 4n) = [\frac{1}{2}N(D,n)^2 - \frac{3}{2}N(D,n) + m] \cdot \quad (13)$$

$$\cdot (4n \cdot N(D,n)^2 + (4n+4) \cdot N(D,n) + 4n) =$$

$$= 2n \cdot N(D,n)^4 - 6n \cdot N(D,n)^3 + 4mn \cdot N(D,n)^2 +$$

$$+ (2n+2) \cdot N(D,n)^3 - (6n+6) \cdot N(D,n)^2 + (4mn+4m) \cdot N(D,n) + 2n \cdot N(D,n)^2 -$$

$$- 6n \cdot N(D,n) + 4mn = 2n \cdot N(D,n)^4 - (4n-2) \cdot N(D,n)^3 + (4mn-4n-6) \cdot N(D,n)^2 +$$

$$+ (4mn+4m-4n) \cdot N(D,n) + 4mn$$

$$\therefore T = \frac{m(m-1)}{2} \cdot (4n \cdot N(D,n) + 3n) + (12) + (13) =$$

$$= (2m^2n - 2mn) \cdot N(D,n) + \frac{3}{2}m^2n - \frac{3}{2}mn + (mn+2n) \cdot N(D,n)^4 +$$

$$+ (-m^2n + mn + m + \frac{23}{2}n + 1) \cdot N(D,n)^3 + (-3m^2n - m^2 + mn - n + 5) \cdot N(D,n)^2 +$$

$$+ (-9m^2n + 8mn - m^2 - 5m + 4n - 2) \cdot N(D,n) - \frac{11}{2}m^2n - 2m^2 - \frac{5}{2}mn + 2m +$$

$$+ 2n \cdot N(D,n)^4 - (4n-2) \cdot N(D,n)^3 + (4mn-4n-6) \cdot N(D,n)^2 +$$

$$+ (4mn+4m-6n) \cdot N(D,n) + 4mn =$$

$$= (mn+4n) \cdot N(D,n)^4 + (-m^2n + mn + m + \frac{15}{2}n + 31) \cdot N(D,n)^3 +$$

$$+ (-3m^2n - m^2 + 5mn - 5n - 1) \cdot N(D,n)^2 +$$

$$+ (-7m^2n + 10mn - m^2 - m - 2n - 2) \cdot N(D,n) +$$

$$+ 4m^2n - 2m^2 + 2m$$

So the complexity of S-polynomial reduction algorithm is

$$T = (mn+4n) \cdot N(D,n)^4 +$$

$$+ (-m^2n + mn + m + \frac{15}{2}n + 31) \cdot N(D,n)^3 + (-3m^2n - m^2 + 5mn - 5n - 1) \cdot N(D,n)^2 +$$

$$+ (-7m^2n + 10mn - m^2 - m - 2n - 2) \cdot N(D,n) + 4m^2n - 2m^2 + 2m$$

This completes the proof. □

### 3.4 Complexity comparison and evaluation

When we put the complexities of above algorithms together and compare, then that is following.

**Table 1. the complexity of S-polynomial reduction sub-algorithm**

| | algorithm | complexity |
|---|---|---|
| 1 | F5-reduction algorithm of F5B | $(2n \cdot N(D,n)^2 + (2n+2) \cdot N(D,n)) \cdot |B_{i-1}|^2 +$ $+ (4n \cdot N(D,n)^2 + 7n \cdot N(D,n)) \cdot |B_{i-1}| +$ $+ (mn+n) \cdot N(D,n)^3 + (mn+m+1) \cdot N(D,n)^2$ |
| 2 | new S-reduction algorithm | $(2n \cdot N(D,n)^2 + 7n \cdot N(D,n)) \cdot |B_{i-1}| +$ $+ (mn+n) \cdot N(D,n)^3 +$ $+ (mn+m+2n+1) \cdot N(D,n)^2 +$ $+ (n+2) \cdot N(D,n).$ |

As you see in the table 1, the complexity of F5-reduction algorithm of F5B is about $N(D,n)^2 \cdot |B_{i-1}|^2$ but the complexity of new S-reduction algorithm that we suggest is about $N(D,n)^2 \cdot |B_{i-1}|$, so our algorithm is very efficient because the deg of $|B_{i-1}|$ is lower as one deg.

As you see in the table 2, the complexity of Buchberger algorithm and F5B is about $n \cdot N(D,n)^5$ but the complexity of new algorithm that we suggest is about $mn \cdot N(D,n)^4$, so our algorithm is very efficient because the deg of $N(D,n)$ is lower as one deg.

**Table 2. the complexity of algorithm for computing Gröbner basis**

| | algorithm | complexity |
|---|---|---|
| 1 | Buchberger algorithm | $\frac{3}{2} n \cdot N(D,n)^5 + N(D,n)^4 + 2mn \cdot N(D,n)^3$ $- m^2 n \cdot N(D,n)^2 - \frac{1}{2} m^2 n \cdot N(D,n)$ |
| 2 | F5B algorithm | $\frac{2n}{3} \cdot N(D,n)^5 + mn \cdot N(D,n)^4 -$ $- m^2 n \cdot N(D,n)^3 + \frac{10}{3} m^3 n \cdot N(D,n)^2 -$ $- \frac{2}{3} m^3 n \cdot N(D,n) - \frac{11}{2} m^2 n$ |
| 3 | New algorithm | $mn \cdot N(D,n)^4 - m^2 n \cdot N(D,n)^3 -$ $- 3m^2 n \cdot N(D,n)^2 - 7m^2 n \cdot N(D,n) +$ $+ 4m^2 n$ |

**Attention:** we know that $m < N(D,n)$ for m and $N(D,n)$.

## 4. CONCLUSIONS
We proposed a pair-selection-method newly to reduce S-polynomial quickly in the algorithms for computing Grobner basis.

## 5. FURTHER STUDY
In the future, we think it is very necessary to research the method for constructing Grobner bases like as mathematical programming. So we are going to make a lot of study to express the F4 algorithm already presented in [2] like as mathematical programming and combine the main idea of F5 with this in the future.

## 6. ACKNOWLEDGMENTS
Authors would like to thank anonymous reviewers' help and advice.